\documentclass[reprint,prl,aps,superscriptaddress,showpacs,amsmath,amssymb]{revtex4-2}
\usepackage{graphicx,float}
\usepackage{dcolumn,tabularx}
\usepackage{bm,color}
\usepackage{hyperref}
\hypersetup{
	colorlinks=true,
	linkcolor=blue,
	citecolor=blue,
	urlcolor=blue,
}
\usepackage{bbding}

\begin{document}
	\title{
		{Demonstrating shareability of multipartite Einstein-Podolsky-Rosen steering}
	}

	\author{Ze-Yan~Hao}
	\affiliation{CAS Key Laboratory of Quantum Information, University of Science and Technology of China, Hefei 230026, People's Republic of China}
	\affiliation{CAS Center For Excellence in Quantum Information and Quantum Physics, University of Science and Technology of China, Hefei 230026, People's Republic of China}
	
	\author{Kai~Sun}
	\email{ksun678@ustc.edu.cn}
	\affiliation{CAS Key Laboratory of Quantum Information, University of Science and Technology of China, Hefei 230026, People's Republic of China}
	\affiliation{CAS Center For Excellence in Quantum Information and Quantum Physics, University of Science and Technology of China, Hefei 230026, People's Republic of China}

	\author{Yan~Wang}
	\affiliation{CAS Key Laboratory of Quantum Information, University of Science and Technology of China, Hefei 230026, People's Republic of China}
	\affiliation{CAS Center For Excellence in Quantum Information and Quantum Physics, University of Science and Technology of China, Hefei 230026, People's Republic of China}
	
	\author{Zheng-Hao~Liu}
	\affiliation{CAS Key Laboratory of Quantum Information, University of Science and Technology of China, Hefei 230026, People's Republic of China}
	\affiliation{CAS Center For Excellence in Quantum Information and Quantum Physics, University of Science and Technology of China, Hefei 230026, People's Republic of China}
	
	\author{Mu~Yang}
	\affiliation{CAS Key Laboratory of Quantum Information, University of Science and Technology of China, Hefei 230026, People's Republic of China}
	\affiliation{CAS Center For Excellence in Quantum Information and Quantum Physics, University of Science and Technology of China, Hefei 230026, People's Republic of China}

	\author{Jin-Shi~Xu}
	\email{jsxu@ustc.edu.cn}
	\affiliation{CAS Key Laboratory of Quantum Information, University of Science and Technology of China, Hefei 230026, People's Republic of China}
	\affiliation{CAS Center For Excellence in Quantum Information and Quantum Physics, University of Science and Technology of China, Hefei 230026, People's Republic of China}
	
	\author{Chuan-Feng~Li}
	\email{cfli@ustc.edu.cn}
	\affiliation{CAS Key Laboratory of Quantum Information, University of Science and Technology of China, Hefei 230026, People's Republic of China}
	\affiliation{CAS Center For Excellence in Quantum Information and Quantum Physics, University of Science and Technology of China, Hefei 230026, People's Republic of China}
	
	\author{Guang-Can~Guo}
	\affiliation{CAS Key Laboratory of Quantum Information, University of Science and Technology of China, Hefei 230026, People's Republic of China}
	\affiliation{CAS Center For Excellence in Quantum Information and Quantum Physics, University of Science and Technology of China, Hefei 230026, People's Republic of China}

	\begin{abstract}
Einstein-Podolsky-Rosen (EPR) steering, a category of quantum nonlocal correlations describing the ability of one observer to influence another party's state via local measurements, is different from both entanglement and Bell nonlocality by possessing an asymmetric property. For multipartite EPR steering, the monogamous situation, where two observers cannot simultaneously steer the state of the third party, has been investigated rigorously both in theory and experiment. In contrast to the monogamous situation, the shareability of EPR steering in reduced subsystems allows the state of one party to be steered by two or more observers and thus reveals more configurations of multipartite EPR steering. However, the experimental implementation of such a kind of shareability has still been absent until now. Here, in an optical experiment, we provide a proof-of-principle demonstration of the shareability of EPR steering  without the constraint of monogamy in a three-qubit system. Moreover, based on the reduced bipartite EPR steering detection results, we verify the genuine three-qubit entanglement results. This work provides a complementary viewpoint for understanding multipartite EPR steering and has potential applications in many quantum information protocols, such as multipartite entanglement detection, quantum cryptography, and the construction of quantum networks.
	\end{abstract}
	
	\maketitle
	
Einstein-Podolsky-Rosen (EPR) steering was first introduced by Schr{\"o}dinger \cite{schrodinger1935discussion} to argue the “action at a distance” paradox in the famous work by Einstein, Podolsky, and Rosen \cite{einstein1935can}. It describes the process in which one observer can  steer another observer’s state through local measurements. Wiseman $et\ al$. developed this concept in 2007 and gave an operational definition \cite{wiseman2007steering}. EPR steering is verified when the assemblage of conditional states of one party $\{\rho_{a|r}\}$, in which the measure direction $r$ is performed by the other party, said Alice, with the output result $a\in \{0,1\}$, cannot be explained via a local hidden state (LHS) model, i.e., the following equation is violated,
\begin{equation}
\rho_{a|r}=\int P(a | r, \lambda) p_{\lambda} \rho_{\lambda}\mathrm{d}\lambda.
\end{equation}
Here $p_\lambda$ parameterized by the hidden variable $\lambda$ is the distribution of LHS $\rho_\lambda$, and $P(a|r, \lambda)$ represents the corresponding probability distribution \cite{cavalcanti2016quantum,uola2020quantum}. While the quantum mechanical prediction of $\rho_{a|r}$ is given by ${\rm Tr_B} \rho_{AB}M_{a|r}$, where $\rho_{AB}$ is the shared state and $M_{a|r}$ is the performed measurement. As a kind of quantum nonlocal correlations, EPR steering, regarded as a one-side device-independent scenario of entanglement verification, lies between quantum entanglement \cite{horodecki2009quantum} and Bell nonlocality \cite{brunner2014bell} hierarchically. Due to its unique directional property, EPR steering indicates an asymmetric manifestation, which further leads to one-way EPR steering, that is Alice can steer Bob but not vice versa \cite{handchen2012observation,bowles2014one,wollmann2016observation,sun2016experimental,xiao2017demonstration}. Since it was reformulated, EPR steering, as well as its application in quantum information tasks, has attracted much attention from the quantum correlation community \cite{branciard2012one,gehring2015implementation,piani2015necessary,sun2018demonstration,vsupic2016self,gheorghiu2017rigidity,gallego2015resource}.

\begin{figure}[t]
		\centering
		\includegraphics[width=0.48\textwidth]{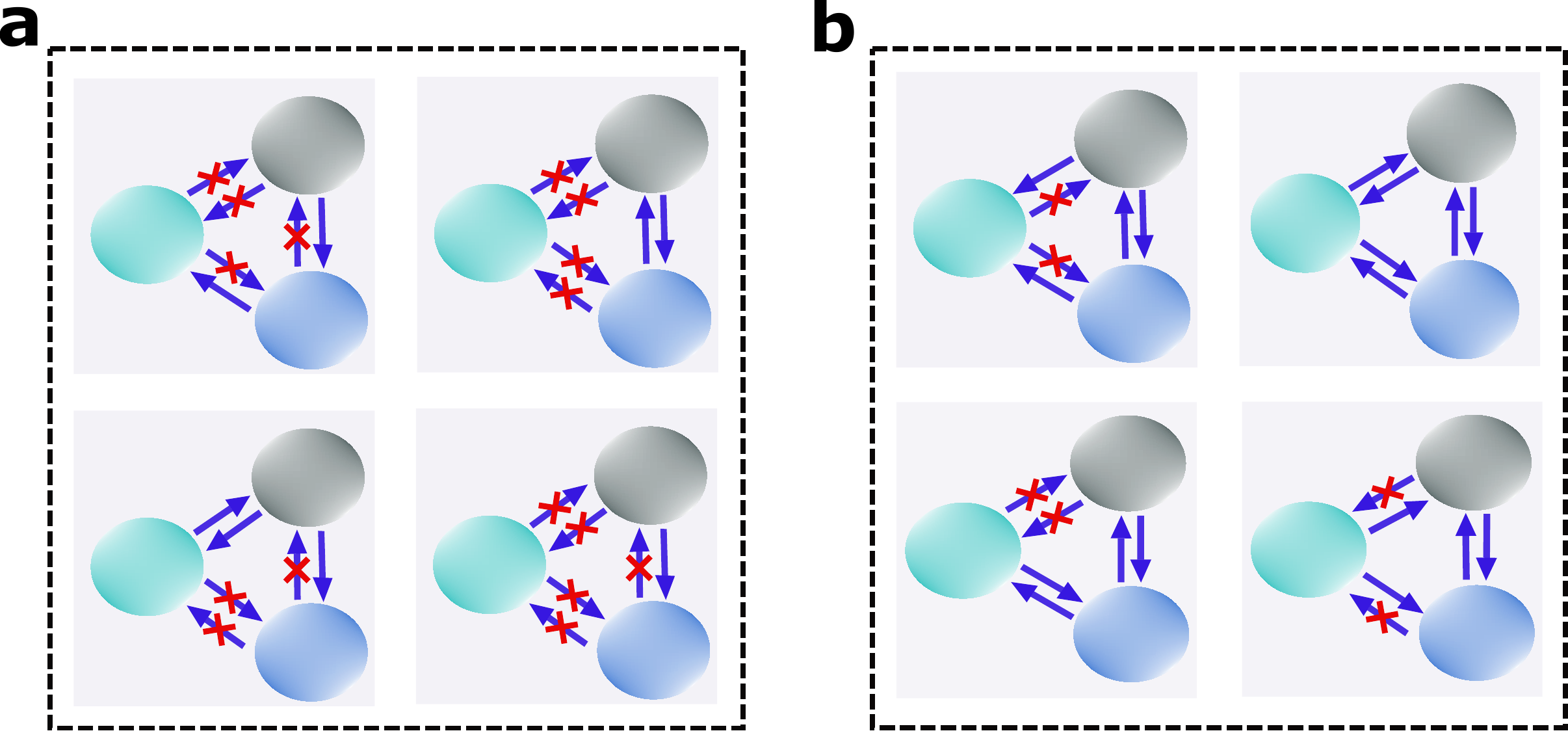}
		\caption{Configurations of the EPR steering correlations shared among three observers. \textbf{a}. The monogamous relations indicate that one observer cannot be steered by the others simultaneously. \textbf{b}. The shareability of EPR steering, in contrast to monogamy, indicates that two observers can simultaneously steer the third observer.
		}\label{p0}
	\end{figure}

In multipartite quantum systems,  EPR steering has been investigated with different approaches, including the one-sided device-independent scenario \cite{cavalcanti2011unified} and the steering correlations between the bipartition \cite{he2011entanglement,he2013genuine,li2015genuine}. As a fundamental feature in multipartite systems, monogamy constraints of EPR steering limit the free distribution of this type of quantum correlation over different subsystems  \cite{reid2013monogamy,PhysRevLett.117.220502,xiang2017multipartite}. This property is analogous to the monogamy of entanglement \cite{coffman2000distributed} and Bell nonlocality \cite{kurzynski2011correlation,cheng2017anisotropic,zhu2019experimental}. Taking a tripartite system (e.g., Alice, Bob, and Charlie) as an illustration, the monogamy of EPR steering refers to the impossibility of Alice and Bob simultaneously steering the state of Charlie \cite{reid2013monogamy} as shown in configurations of Fig. \ref{p0}\textbf{a}. Monogamous relations of EPR steering have drawn much attention and been studied in both theory and experiment within multipartite and high-dimensional quantum systems \cite{armstrong2015multipartite,deng2017demonstration,deng2018quantification,wang2020deterministic,cai2020versatile,xiang2020monogamy,zhang2019experimental,milne2014quantum}.

On the other hand, there is an attractive attempt to break monogamous relations to reveal more configurations of multipartite EPR steering \cite{uola2020quantum}. It has been shown that monogamy constraints could be removed by increasing the number of measurement settings \cite{reid2013monogamy,paul2020shareability}. These efforts uncover a property opposite to monogamy, referred to as the shareability of EPR steering over subsystems. In such a case, Alice and Bob can steer Charlie simultaneously as shown in Fig. \ref{p0}\textbf{b} \cite{paul2020shareability}.
Identification of EPR steering shareability provides a comprehensive insight to understand EPR steering distributed over multiple parties and enriches the scenarios of quantum tasks based on multipartite steering, such as quantum internet \cite{kimble2008quantum,wang2020deterministic,PhysRevLett.127.170405}, quantum authentication \cite{huang2019securing,mondal2019authentication}, EPR steering swapping \cite{wang2017einstein}, and the information loss paradox of the black hole \cite{wang2018monogamy}. However, an experimental demonstration of EPR steering shareability is still absent.

In this work, based on a three-qubit system consisting of the three degrees of freedom (DOF) of a single photon, namely, the polarization, path, and orbital angular momentum, different configurations of EPR steering shareability are experimentally demonstrated. Considering the directional property of EPR steering, these relations are observed by exploiting the uncertainty relation criterion with three measurement settings \cite{PhysRevA.93.012108}. The experimental result, which demonstrates Alice can be steered by Bob and Charlie simultaneously for the first time, qualifies as a proof-of-principle experimental observation of EPR steering shareability without the constraint of monogamy. Moreover, the detection of EPR steering shareability  facilitates the verification of genuine three-qubit entanglement. Our results can be a significant step forward to extending the understanding of multipartite relationships and have potential applications in quantum information protocols  \cite{mardari2021erase,uola2020quantum}.

The shareability relations of EPR steering explicitly proposed in Ref. \cite{paul2020shareability} employ the violation of linear steering inequality. However, since this criterion cannot reflect the directional property of EPR steering, we exploit a stronger criterion based on uncertainty relations \cite{PhysRevA.93.012108} to observe more EPR steering shareability configurations.
Concretely, Alice can steer Bob if the inequality
	\begin{equation}
	P_{AB}=	\sum_{i} \delta^{2}\left(\alpha_{i} A_{i}+B_{i}\right) \geq \min _{\rho_{B}} \sum_{i} \delta^{2}\left(B_{i}\right)\label{un}
	\end{equation}
is violated, where $\delta$ denotes the variance of the measurement outcomes and $\alpha_{i}=-\frac{C\left(A_{i}, B_{i}\right)}{\delta^{2}\left(A_{i}\right)}$ in which $C\left(A_{i}, B_{i}\right)=\left\langle A_{i} B_{i}\right\rangle-\left\langle A_{i}\right\rangle\left\langle B_{i}\right\rangle$. For a three-qubit system (Alice, Bob, and Charlie), in the case where the number of measurement settings $n=2$, the monogamous relation is valid. However, by increasing $n$ to 3, the monogamy violations are possible, which means that more shareability configurations of EPR steering can be observed. Here, three measurement settings are chosen as $\left\{\sigma_{x},\ \sigma_{y},\ \sigma_{z}\right\}$. We can confirm that $\min _{\rho_{A}} \sum_{i} \delta^{2}\left(A_{i}\right)=\min _{\rho_{A}} \sum_{i=x, y, z}\left(1-\left\langle\sigma_{i}\right\rangle^{2}\right)=2$, similarly,
	$
	\min _{\rho_{B}} \sum_{i} \delta^{2}\left(B_{i}\right)=\min _{\rho_{C}} \sum_{i} \delta^{2}\left(C_{i}\right)=2$. The parameter $P_{AB}<2$ violates the inequality (\ref{un}) and indicates that Bob can be steered by Alice. Many contents of EPR steering shareability relations emerge upon exploiting this criterion. Taking the
	W-like states \cite{dur2000three} as an illustration,
	\begin{equation}
	\begin{aligned}
	\left|\psi_{A B C}\right\rangle =&\alpha\left| 0\right\rangle\left| 0\right\rangle\left| 1\right\rangle +\beta\left| 0\right\rangle\left| 1\right\rangle\left| 0\right\rangle +\gamma\left| 1\right\rangle\left| 0\right\rangle\left| 0\right\rangle, \label{W}
	\end{aligned}
	\end{equation}
where $|\alpha|^2+|\beta|^2+|\gamma|^2=1$. For $\alpha=0.2$, $ \beta=0.4$, and $\gamma =\sqrt{0.8}$, the parameters $P_{BA},\ P_{AB}$, and $ P_{AC}$ violate the steering inequality \eqref{un}, which means that monogamy can be observed when no one can be steered by the others at the same time. For $\alpha=1/2$, $ \beta=1/2$, and $\gamma =1/\sqrt{2}$, we can obtain $P_{B A}=P_{C A}<2$, which indicates that Alice can be steered by Bob and Charlie simultaneously. In this assessment, more  EPR steering configurations can be observed. In particular, for some states, Alice, Bob, and Charlie can steer each other in the case where all steering parameters are less than 2 \cite{reid2013monogamy}. For instance, the steering parameters of state $\left|W\right\rangle=(|001\rangle+|010\rangle+|100\rangle)/\sqrt{3} $ all have the theoretical values of $16/9<2$. More details regarding shareability relations of EPR steering are shown in the Supplementary Material (SM) \cite{hhh}. 	
Furthermore, based on the detection of EPR steering shareability relations, genuine multipartite entanglement can be verified when the monogamous relationship vanishes \cite{paul2020shareability}.

		\begin{figure*}[t]
		\centering
		\includegraphics[width=0.8\textwidth]{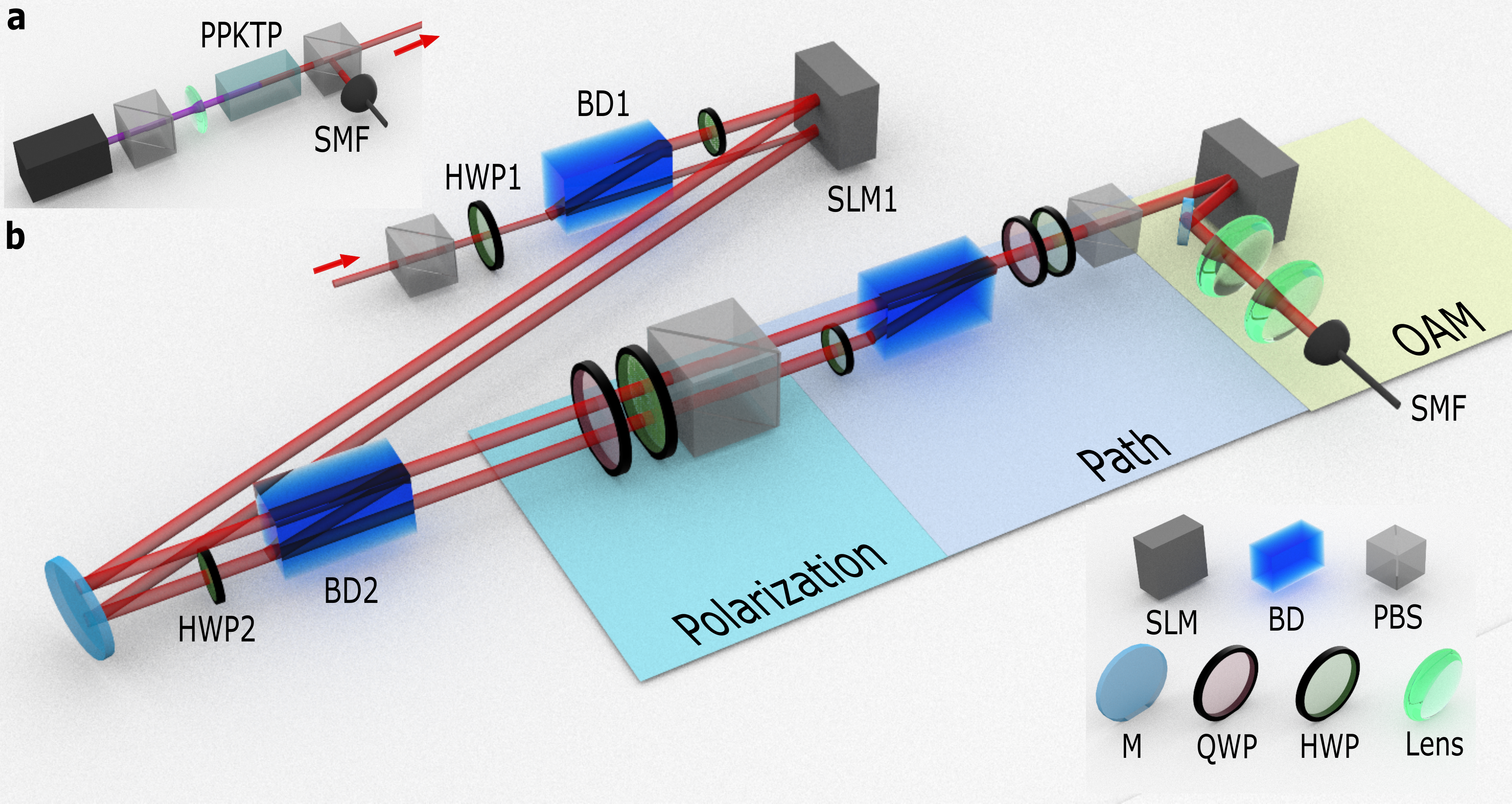}
		\caption{
			Illustration of the process of preparation and measurement of three-qubit states.  \textbf{a}.  A heralded single-photon source generates a pair of photons via a PPKTP crystal. \textbf{b}. The main optical system. The initial photon is divided into two optical paths through the  half wave plate (HWP1) and the beam displacer (BD1). The first spatial light modulator (SLM1) then generates an orbital angular momentum (OAM) via the phase-only hologram. The GHZ-like states can be prepared when the angle of the HWP2 is set to
			$45^{\circ}$. The BD2 is positioned to generate W-like states and adjust their parameters in Eq. \eqref{W} by changing HWP1 and HWP2. The measurement apparatus is composed of three independent parts that can achieve separate projective measurements of states with different degrees of freedom  (polarization, path, and OAM).
		}\label{p1}
	\end{figure*} 	
	
{\it Experimental setup.}	
As shown in Fig. \ref{p1}\textbf{a}, the heralded single photons \cite{eisaman2011invited} is generated from a 20 mm long  periodically poled KTiOPO$_4$ (PPKTP) crystal which is pumped by a 404-nm continuous-wave laser \cite{fedrizzi2007wavelength}. The initial photon polarization state is prepared to be $\mu \left| H\right\rangle +\nu\left| V\right\rangle $ ($|\mu|^2+|\nu|^2=1$), in which $ \left| H\right\rangle$ and $\left| V\right\rangle$ correspond to the horizontal and vertical polarization bases, respectively. A half-wave plate (HWP1) is used to adjust $\mu$ and $\nu$. Using a  beam displacer (BD1), which can split the input into two orthogonally polarized beams, the $\left| H\right\rangle$ and $\left| V\right\rangle$  states are separated into different paths, $\left| H\right\rangle$ for the down-path and $\left| V\right\rangle$ for the up-path.  The first DOF can then be represented by the orthogonal path-basis $\left| U\right\rangle$ (up-path) and $ \left| D\right\rangle $ (down-path). After passing through the BD1, the initial state becomes $\mu \left| D\right\rangle +\nu\left| U\right\rangle $. The second DOF  is given by orbital angular momentum (OAM) generated via the spatial light modulator (SLM1) \cite{bolduc2013exact}. The upside hologram exhibits a $\left| -l\right\rangle $ grating, whereas the downside hologram generates a $\left| +l\right\rangle $ grating, where $\left| \pm l\right\rangle $ correspond to the Laguerre-Gaussian (LG) modes and represent the states with orbital angular momentum $\pm l\hbar$. In this experiment, $l$ is set to be 2. Since the SLM only works for the $ \left| H\right\rangle$ polarization, a $45^{\circ}$ HWP is used in front of the SLM1 to turn the polarization of the up-path into  $ \left| H\right\rangle$. After passing through the SLM1, the state evolves to $\mu\left| D\right\rangle \left| +l\right\rangle +\nu\left| U\right\rangle\left| -l\right\rangle$. The HWP2 after the SLM1 is then used  to transform the down-path polarization into $\left| V\right\rangle$. In this way, a Greenberger-Horne-Zeilinger (GHZ) like state, namely, $\left| G\right\rangle  =\mu\left| V\right\rangle\left| D\right\rangle\left| +l\right\rangle +\nu\left| H\right\rangle\left| U\right\rangle\left| -l\right\rangle$ could be prepared using the three DOF of the single photon which is equivalent to the states generated with three photons \cite{vitelli2013joining}. To obtain the target W-like states, another beam displacer (BD2) is used to generate more components. After passing through BD2, the path of the $\left| H\right\rangle $ polarization remains unchanged, and the path of the $\left| V\right\rangle $ polarization deviates into a higher path ($\left| V\right\rangle\left| D\right\rangle \to \left| V\right\rangle\left| U\right\rangle$). By rotating the angle $\theta$ of the HWP2, the state $\left| V\right\rangle\left| D\right\rangle$ could be prepared to be $\operatorname{sin}2\theta\left| V\right\rangle\left| D\right\rangle+\operatorname{cos}2\theta \left| H\right\rangle\left| D\right\rangle$. Thus, the state
$\left| V\right\rangle\left| D\right\rangle\left| +l\right\rangle $
becomes $\operatorname{cos}2\theta\left| H\right\rangle\left| D\right\rangle\left| +l\right\rangle
+\operatorname{sin}2\theta\left| V\right\rangle\left| U\right\rangle\left| +l\right\rangle$, whereas the states $\left| H\right\rangle\left| U\right\rangle\left| -l\right\rangle$ does not change.
With encoding $\left| H\right\rangle \to\left| 0\right\rangle_s$, $\left| V\right\rangle \to\left| 1\right\rangle _s$, $\left| U\right\rangle \to\left| 0\right\rangle_p$, $\left| D\right\rangle \to\left| 1\right\rangle _p$, $\left| +l\right\rangle \to\left| 0\right\rangle_m$, and $\left| -l\right\rangle \to\left| 1\right\rangle _m $, the  W-like states in Eq. \eqref{W} are prepared in which the parameters $\alpha$, $\beta$, and $\gamma$ are determined by the angles of HWP1 and HWP2.

The measurement process can be divided into three independent projective measurements, namely, polarization, path, and OAM analyzer \cite{wang201818}. The unit of a polarization analyzer is composed of a quarter-wave plate (QWP), an HWP, and a polarization beam splitter (PBS). The path measurement contains an HWP, a BD, and a polarization analyzer, in which the HWP and the BD are used to convert path information into polarization information. After passing through the BD, the $\left| U\right\rangle $ path changes into the $\left| H\right\rangle $, and the $\left| D\right\rangle $ path is converted into $\left| V\right\rangle $. Therefore, the polarization analyzer is used after the BD to realize a projective path measurement. The OAM measurement consists of an SLM loading phase-only hologram \cite{bolduc2013exact} and a single mode fiber (SMF). Different projective measurements of the OAM qubit can be achieved by transforming target states (like $(\left| +l\right\rangle +\left| -l\right\rangle)/\sqrt{2} $) into the $l=0$ mode with different holograms generated by the SLM. The SMF is then used to couple the $l=0$ mode and filter out the other modes. Before being detected by the single-photon detectors, the photons in Fig. \ref{p1}\textbf{a} and Fig. \ref{p1}\textbf{b} are filtered through interference filters with a bandwidth of 3 nm, and the resulting signals are then sent to coincidence counting.

	\begin{figure*}[ht]
		\centering
		\includegraphics[width=0.94\textwidth]{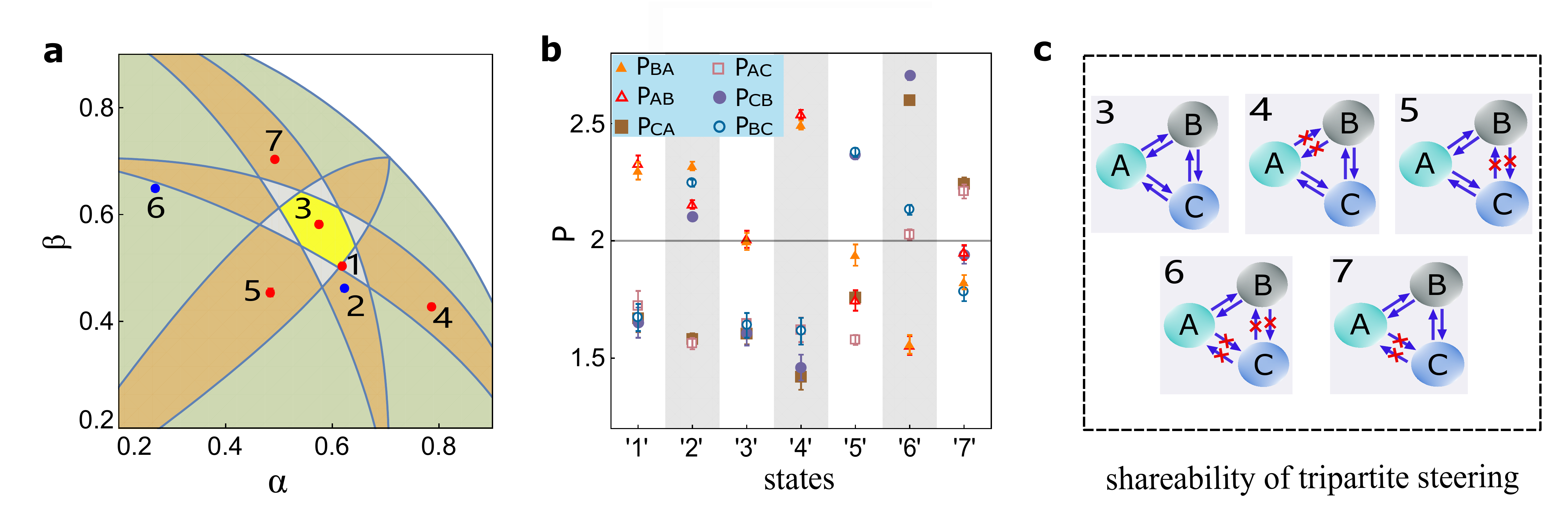}
		\caption{ Experimental results. \textbf{a}. The  horizontal and vertical coordinates are $\alpha$ and $\beta$ in Eq. \eqref{W}. The background of this figure is divided into different areas to indicate different relationships. The red experimental dots demonstrate the EPR steering shareability relations existing without the monogamy constraint. The states exhibiting monogamous relations are marked by blue dots. The experimentally prepared three-qubit states are distinguished by the numbers marked beside them.
			\textbf{b} indicates the results obtained based on the uncertainty relation criterion. Different parameters are labeled by different colors and shapes. Alice can steer Bob when the steering parameter $P_{AB}<2$, whereas $P_{BA}<2$ indicates that Bob can steer Alice. \textbf{c}. The different shareability relations of tripartite EPR steering correspond to the states shown in \textbf{a}.   }\label{bb2}
	\end{figure*}

{\it Experimental results.}	
Seven W-like states are experimentally prepared as in Eq. \eqref{W} with an average fidelity  $F=\operatorname{Tr}\left(\sqrt{\sqrt{\rho_{\mathrm{th}}} \rho_{\mathrm{ex}} \sqrt{\rho_{\mathrm{th}}}}\right)$ of 0.960(3) \cite{nery2020distillation}. Here, $\rho_{\mathrm{ex}}$ is obtained through the experimental results, whereas $\rho_{\mathrm{th}}$ represents the ideal theoretical state. More details are provided in the SM \cite{hhh}.
The parameter $P_{BA}<2$ indicates that Alice can be steered by Bob. The red dots in Fig. \ref{bb2}\textbf{a} demonstrate the shareability of EPR steering without the constraint of monogamy in cases with three measurement settings. For example, for state ``4", the steering parameters $P_{AC}=1.62(5)$ and $P_{BC}=1.61(5)$ shown in Fig. \ref{bb2}\textbf{b} demonstrate that Charlie can be steered by Alice and Bob. This property is similar to that of state ``1". For state ``5", the steering parameters $P_{BA}=1.94(4)$ and $P_{CA}=1.75(3)$ indicate that Alice can be steered by Bob and Charlie, and for state ``7", the same analysis derives the corresponding shareability relations shown in Fig. \ref{bb2}\textbf{c}. In this way, the shareability relations transcending the constraint of EPR steering monogamy are verified via the uncertain relation criterion. The blue dots in Fig. \ref{bb2}\textbf{a} represent monogamy even in the case with three measurement settings. For example, state ``6" with the parameters $P_{BA}=1.56(4)$, $P_{AB}=1.55(4)$, $P_{CA}=2.60(2)$, $P_{AC}=2.03(2)$, $P_{CB}=2.70(2)$, and $P_{BC}=2.13(2)$ suggests that Alice can only steer Bob, while Bob can only steer Alice.
Moreover, in the yellow area of Fig. \ref{bb2}\textbf{a}, Alice, Bob, and Charlie can be steered by each other simultaneously. For state ``3" with parameters $P_{BA}=1.99(3),\ P_{AB}=2.00(3),\ P_{CA}=1.60(4),\ P_{AC}=1.65(5), \ P_{CB}=1.61(5)$, and $ P_{BC}=1.64(4)$, violation of the  monogamy relation by the shareability of EPR steering can then be presented.  The shareability of two-way correlations in all bipartite states also indicates a stronger multipartite EPR steering \cite{reid2013monogamy}.
The error bars in the figures are handled by Poisson counting statistics. We also employ another criterion that can detect the shareability of EPR steering via only the tomographic measurements of reduced single-qubit states (RSQSs) \cite{paul2020shareability}. Compared to the uncertainty relation criterion, this method needs fewer projectors but sacrifices some valid ranges. More details are introduced in the SM \cite{hhh}.
	
	\begin{table}[!t]
		\caption{ Results of the detected of  EPR steering shareability relations (SR) and witness detection.  The ``Y" and ``N" of ``SR" indicate whether genuine entanglement has been detected by shareability relations.  The negative values of the witness represent a double confirmation of the genuine entanglement.   }
		\begin{ruledtabular}
			\begin{tabular}{ccccccc}
				& state &   SR &  witness  \\ \hline
				&   1 \  &  Y \    & -0.27(1)  \\
				&   2 \  &  N \    & -0.23(1)  \\
				&   3 \  &  Y \    & -0.26(1)  \\
				&   4  \ &  Y \    & -0.14(1) \\
				&   5  \ &  Y \    & -0.21(1)  \\
				&   6 \  &  N \    & -0.15(1)  \\
				&   7  \ &  Y \    & -0.23(1)  \\
				
			\end{tabular}
		\end{ruledtabular}\label{T1}
	\end{table}
	
Furthermore, the verification results of the EPR steering shareability relations are used to test whether the states are genuinely entangled \cite{paul2020shareability}. As a comparison, the three-qubit witness \cite{acin2001classification} is employed for double verification. The witness
$\mathcal{W}=\frac{2}{3} \mathcal{I}-P_{W}\label{wit}$ is used, where $P_{W}$ is the projector of $\left| W\right\rangle =(\left| 001\right\rangle +\left| 010\right\rangle +\left| 100\right\rangle)/\sqrt{3}$, and $ \mathcal{I}$ is the identity matrix. The result that $\operatorname{Tr}(\rho \mathcal{W})<0$ indicates a correlation of genuine entanglement. The experimental results are presented in Table \ref{T1}, where conclusions regarding the detection of  EPR steering shareability relations and witness detection are listed.  All the states have negative witness values, whereas only the states represented by red dots in Fig. \ref{bb2}\textbf{a} can be verified as being genuinely entangled since they support the EPR steering shared among the three observers.  The criterion of Eq. \eqref{un} leads to the fact that several W-like states, which are genuinely three-qubit entangled states that have the property of EPR steering being shared only between two observers, such as states ``2" and ``6", cannot be detected by the proposed method. This indicates that the verification of EPR steering shareability relations is a sufficient and unnecessary method to test for genuine three-qubit entanglement.
	
In summary, in contrast with previous works that demonstrated the monogamous relations of multipartite EPR steering, in this work, we observe the shareability relations of EPR steering, which are separate from monogamy with multisetting measurements. Based on a multimode interference setup capable of simultaneously manipulating multiple DOFs of photons with high precision, we experimentally verify the shareability relations of EPR steering in reduced subsystems by exploiting a series of W-like states and the criterion of uncertainty relations. We also employ the shareability of multipartite EPR steering to detect genuine tripartite
entanglement.

Extending to a quantum system consisting of $N$ particles, the configurations of EPR steering shareability become more complex, and the scenario that $N-1$ particles simultaneously steer the rest one may be difficult to realize with the increase of $N$ (even to infinite). Taking the state $\left|W_{N}\right\rangle=1 / \sqrt{N}(|00 \cdots 1\rangle+|0 \cdots 10\rangle+\cdots+|10 \cdots 0\rangle)$ \cite{dur2000three} as an example, when $N$ increases from 3 to 4, all the reduced steering parameters are obtained as $S=13/6>2$ (contrasted with $S=16/9<2$ for the tripartite $W$ state in this work), which cannot demonstrate the EPR steering steerability between any reduced bipartite states, and the state is still entangled. This property is different from the entanglement shareability where only separable states are infinite shareable \cite{seevinck2010monogamy,PhysRevLett.117.060501}.

Our results contribute to a significant step forward in the study of multipartite systems. The exploitation of multisetting scenarios provides a deeper understanding of the steerability shared among reduced subsystems. It would be interesting to extend the proposed method to more complex multipartite systems and observe the EPR steering shareability relations therein. The results of this work provide a valuable method for realizing multipartite genuine entanglement testing \cite{knips2016multipartite}. Since monogamy implies security limits on quantum cryptography \cite{huang2019securing}, our results may provide a basis for more applications of cryptographic protocols based on  EPR steering. Furthermore, as this work demonstrates the abundant configurations of EPR steering shared in a multipartite system, we hope that it can be helpful for building a future multipartite EPR steering network \cite{PhysRevLett.127.170405}.
	
	\begin{acknowledgments}
	We thank Yu Xiang and Qiongyi He for the useful discussion.
	This work was supported by
	National Key Research and Development Program of China (Grants
	Nos.\,2016YFA0302700, 2017YFA0304100),
	the National Natural Science Foundation of China (Grants
	Nos.\,61725504, U19A2075,
	61805227, 61975195,
	11774335, and 11821404),
	Key Research Program of Frontier Sciences, CAS (Grant No.\,QYZDY-SSW-SLH003),
	Science Foundation of the CAS (Grant No.\,ZDRW-XH-2019-1),
	the Fundamental Research Funds for the Central Universities (Grant No.\,WK2470000026, No.\,WK2030380017),
	Anhui Initiative in Quantum Information Technologies (Grants No.\,AHY020100, and No.\,AHY060300).
\end{acknowledgments}

\end{document}